\documentclass[twocolumn,english,reprint,linenumbers]{revtex4}
\usepackage[T1]{fontenc}
\usepackage[latin9]{inputenc}
\usepackage{textcomp}
\usepackage{amsmath}
\usepackage{amssymb}
\usepackage{graphicx}

\makeatletter
\@ifundefined{textcolor}{}
{%
 \definecolor{BLACK}{gray}{0}
 \definecolor{WHITE}{gray}{1}
 \definecolor{RED}{rgb}{1,0,0}
 \definecolor{GREEN}{rgb}{0,1,0}
 \definecolor{BLUE}{rgb}{0,0,1}
 \definecolor{CYAN}{cmyk}{1,0,0,0}
 \definecolor{MAGENTA}{cmyk}{0,1,0,0}
 \definecolor{YELLOW}{cmyk}{0,0,1,0}
 }

\usepackage{babel}

\usepackage{babel}

\makeatother

\usepackage{babel}
\begin{document}

\title{Optimization of indirect magnetoelectric effect in thin-film/substrate/piezoelectric-actuator
heterostructure using polymer substrate}

\author{Mouhamadou Gueye}

\email{mouhamadou.gueye@lspm.cnrs.fr}

\selectlanguage{english}%

\author{Fatih Zighem}

\email{zighem@univ-paris13.fr}

\selectlanguage{english}%

\author{Damien Faurie, Mohamed Belmeguenai and Silvana Mercone }

\affiliation{LSPM-CNRS, Université Paris XIII, Sorbonne Paris Cité, 93430 Villetaneuse,
France}

\date{July 17$^{th}$ 2014 }
\begin{abstract}
Indirect magnetoelectric effect has been studied in magnetostrictive-film/substrate/piezoelectric-actuator
heterostructures. Two different substrates have been employed: a flexible
substrate (Young's modulus of 4 GPa) and a rigid one (Young's modulus
of 180 GPa). A clear optimization of the indirect magnetoelectric
coupling, studied by micro-strip ferromagnetic resonance, has been
highlighted when using the polymer substrate. However, in contrast
to the rigid substrate, the flexible substrate also leads to an \textit{a
priori} undesirable and huge uniaxial anisotropy which seems to be
related to a non equibiaxial residual stress inside the magnetostrictive
film. The {}``strong'' amplitude of this non equibiaxiallity is
due to the large Young's modulus mismatch between the polymer and
the magnetostrictive film which leads to a slight curvature along
a given direction during the elaboration process and thus to a large
magnetoelastic anisotropy.
\end{abstract}
\maketitle
Prospect of local magnetization control using an electric field or
a pure voltage, for low power and ultrafast new electronics has resulted
in an amount of new research domains mainly focused on artificial
engineered materials\cite{Nan2011_Adv_Mat,Ramesh2010_Adv_Mat,Martins2013_Adv_Func_Mater,Lecoeur2011}.
Artificial magnetoelectric systems seems to be a promising route for
such control. The easiest artificial architecture for a magnetization
voltage control in those kinds of materials appears to be the piezoelectric/magnetostrictive
bilayers presenting a good strain-mediated coupling at the interface
\cite{Lecoeur2011,Liu2010,Li2012,Lou2009,Wang2012,Pettiford2008,Zighem_JAP2013,Brandlmaier2009,Brandlmaier2011}.
However, in these systems, clamping effects due to the substrate limits
the strain applicable by the piezoelectric environment to the magnetization
and therefore reduces perspective on applications. Indeed, in this
kind of bilayers, a significant magnetoelectric coupling is obtained
only in the presence of non negligible in-plane stresses in the magnetic
media. Therefore, more the elastic strains are transferred at the
interfaces of such systems, more this indirect magnetoelectric effect
is optimized. Generally, the magnetic thin films are first deposited
on a thick substrate ($\sim$ hundreds of microns), which is then
cemented on a piezoelectric actuator \cite{Pettiford2008,Zighem_JAP2013,Brandlmaier2008_PRB,Brandlmaier2008_PRB_Bis}.
This process limits the desired phenomenon, especially when the substrate
is stiff such as for commonly used wafer (Si, GaAs, ...). This often
reported limitation\cite{Brandlmaier2008_PRB_Bis} (a few ten percents
of losses in best cases) can be avoided by depositing the magnetic
thin film on a compliant substrate such as polyimides \cite{Zighem_JAP2013}
that are more and more used in flexible spintronics\cite{Pinto2014}.

\begin{figure}
\includegraphics[bb=45bp 300bp 775bp 580bp,clip,width=8.5cm]{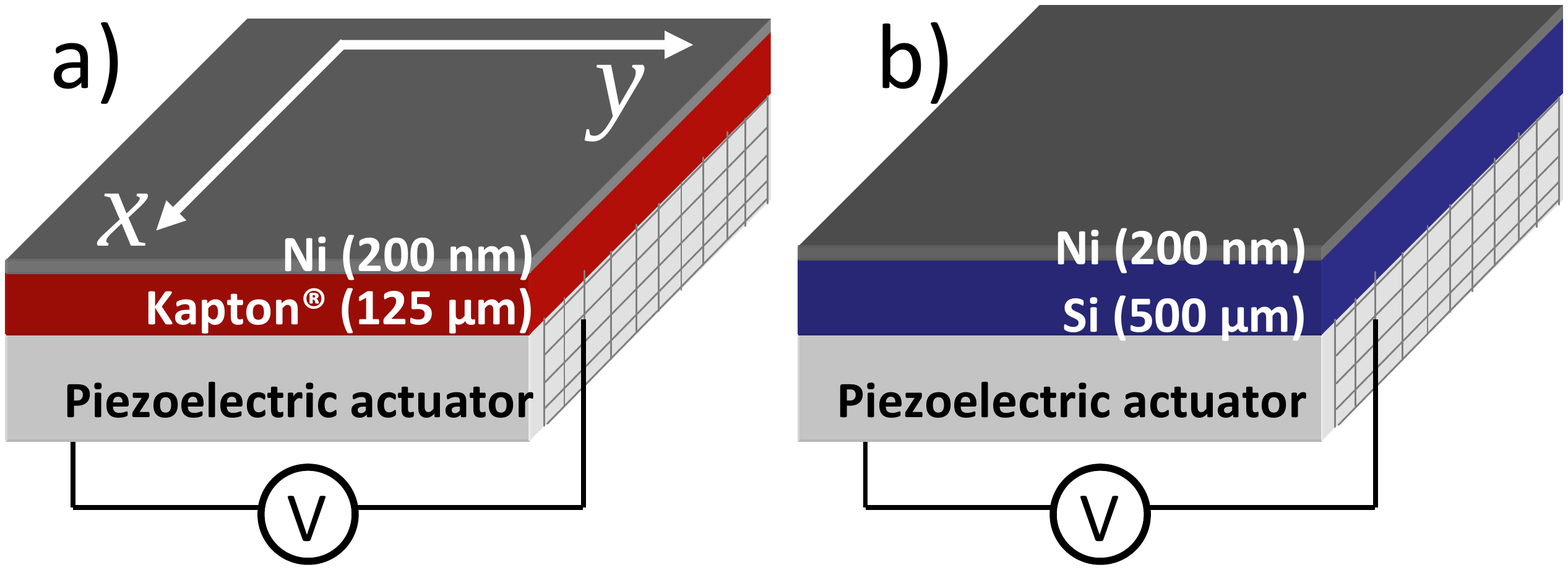}

\caption{Schematics of the two studied heterostructures. The only difference
comes from the substrate (either flexible (a) or rigid (b)) on which
the Ni thin film has been deposited.}

\label{Fig_Sketch_Structures}
\end{figure}

The present study presents a quantitative comparison of the obtained
effective MagnetoElectric (ME) coupling in two different artificial
magnetoelectric heterostructures characterized by the presence of
a flexible or a rigid substrate between a ferromagnetic film and a
piezoelectric actuator. The effective ME coupling will be deduced
from \textit{in situ} MicroStrip FerroMagnetic Resonance (MS-FMR).
The artificial ME heterostructures are composed of a 200 nm Nickel
film deposited by radio frequency sputtering either on a rigid substrate
(Silicon of thickness 500 $\mu$m) or on a flexible one (Kapton\textregistered{}
of thickness 125 $\mu$m) and then glued onto a piezoelectric actuator
as presented on Figure \ref{Fig_Sketch_Structures}. It is worth mentioning
that Ni material has been chosen because of its well-known negative
effective magnetostriction coefficients at saturation even in a polycrystalline
film with no preferred orientations, which is closely the situation
for both systems. In this condition, only one magnetostriction coefficient
at saturation ($\lambda$) is sufficient to characterize the magnetoelastic
anisotropy. Moreover, the two substrates have been chosen because
of their Young's modulus ($\sim$4 GPa for Kapton\textregistered{}
as compared to $\sim$180 GPa for Si).

\begin{figure}
\includegraphics[bb=30bp 130bp 650bp 595bp,clip,width=8cm]{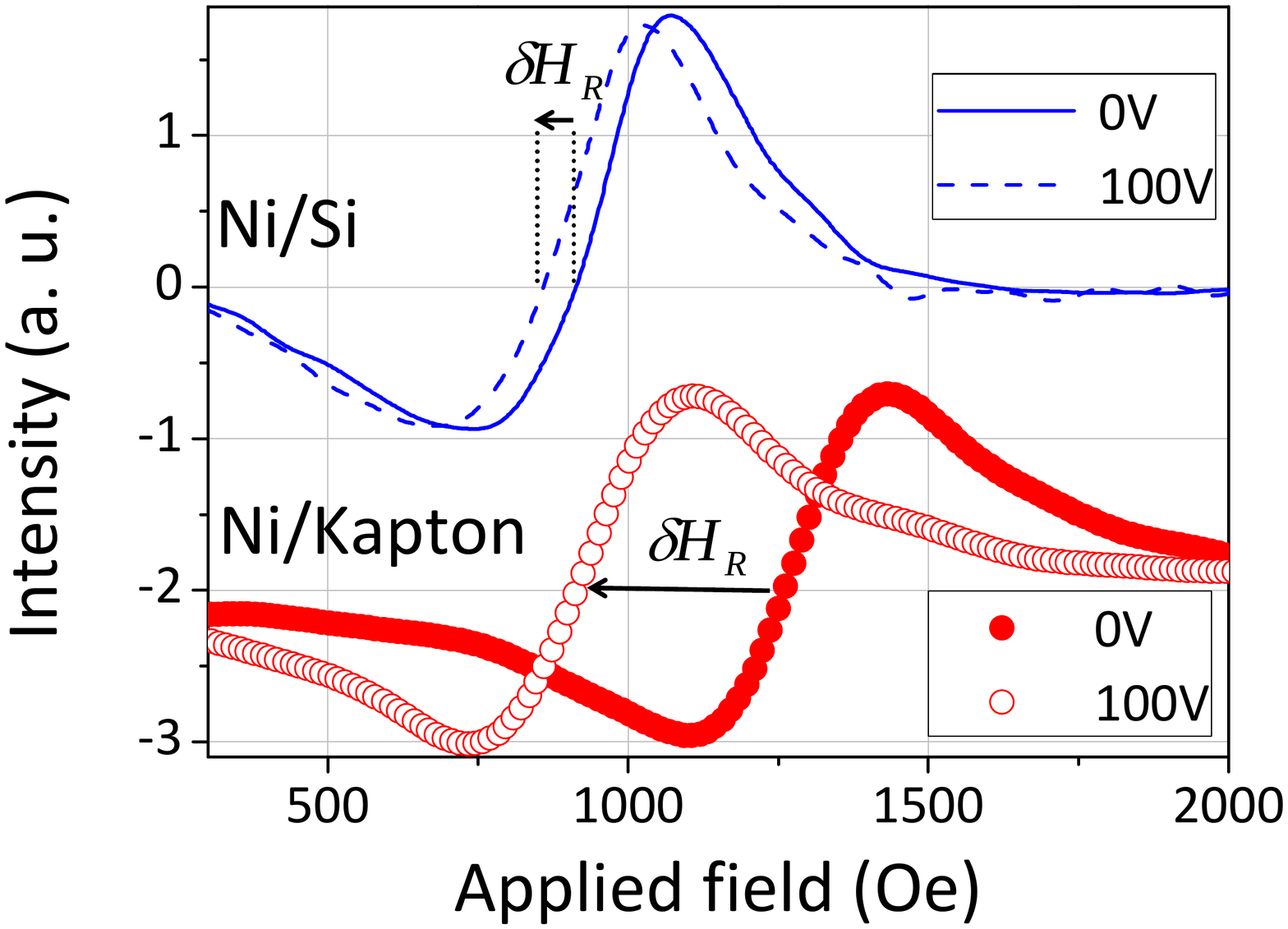}

\caption{Experimental spectra recorded at 8 GHz for an in-plane magnetic field
at 90 degree with respect to the main positive strain axis of the
actuator (along $y$-axis). The largest shift of the resonance is
clearly put into evidence for the Ni film deposited onto Kapton\textregistered{}.}

\label{Fig_FMR_Spectra}
\end{figure}

MS-FMR experiments have been performed at a fixed frequency by sweeping
the applied magnetic field from 0 to 3 kOe \cite{Belmeguenai2013_PRB}.
The microwave driven frequency was fixed at 8 GHz in all the presented
experiments, where the resonance fields are sufficiently higher than
the in-plane magnetic anisotropies fields present in the structures,
avoiding undesirable magnetic and magnetoelastic hysteresis effects.
The \textit{in situ }applied voltage experiments have been done by
varying the external voltage from 0V to 100V and back to 0V with step
of around 5 V. Figure \ref{Fig_FMR_Spectra} shows typical MS-FMR
spectra for two different applied voltages (0 and 100V). The magnetic
field is applied along $y$ direction which corresponds to a negative
induced in-plane strains. Indeed, in the studied range of applied
voltage {[}0-100V{]}, the piezoelectric actuator presents a main positive
in-plane strain along $x$ direction ($\varepsilon_{xx}>0$) and a
smaller negative one along $y$ direction ($\varepsilon_{yy}<0$ with
$\varepsilon_{xx}\sim-2\varepsilon_{yy}$) \cite{Zighem_JAP2013}.
We observed that in both structures the resonance field decreases
as function of the applied voltage. This behavior is consistent with
the negative magnetostriction coefficient at saturation of Ni material
($\lambda<0$). Indeed, in first approximation, the magnetoelastic
anisotropy can be viewed as a voltage-induced uniaxial magnetoelastic
anisotropy field $\vec{H}_{ME}(V)$ along $y$ direction. Thus, since
the \textit{in situ} MS-FMR experiments are performed along $y$,
this $y$-axis will be easier (when increasing the applied voltage)
for the magnetization direction leading to lower values of the resonance
field. Figure \ref{Fig_FMR_Spectra} shows a clear shift of the resonance
field $\delta H_{R}$ (defined as $\delta H_{R}=H_{R}(0)-H_{R}(V)$
) between 0 V and 100 V for both heterostructures: it is close to
50 Oe for Si substrate and reaches a value around 350 Oe when a flexible
substrate is used.

\begin{figure}
\includegraphics[bb=20bp 60bp 700bp 595bp,clip,width=8cm]{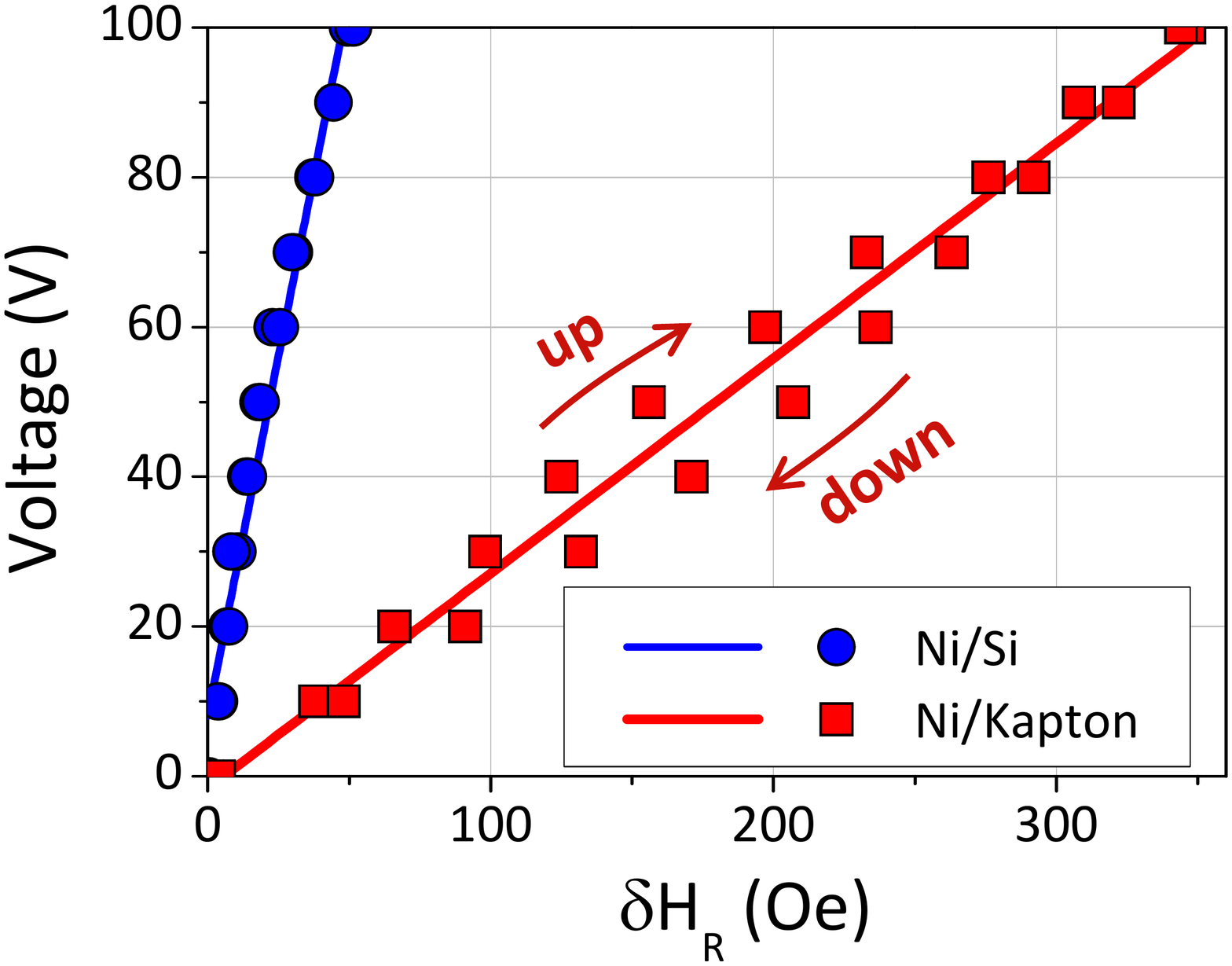}

\caption{Resonance field shift ($\delta H_{R}$ ) variations as function of
the applied voltage for the two studied heterostructures. The lines
correspond to linear fits of the experimental data. Arrows refer to
up and down applied voltage sweeps (0 V to 100 V and 100 V to 0 V,
respectively).}

\label{Fig_Resonance_Field_Shift}
\end{figure}

The complete $\delta H_{R}$ variations as function of the applied
voltage are reported in Figure \ref{Fig_Resonance_Field_Shift} where
circles (resp. squares) refer to results obtained for Si (resp. Kapton\textregistered{})
substrate. A non linear and hysteretic variation is put into evidence
when using Kapton\textregistered{} as a substrate whereas an almost
linear dependence is found in the second structure indicating a difference
of the strain transmission efficiency. This non linear and hysteretic
behavior is due to the intrinsic properties of the ferroelectric material
used during piezoelectric actuator fabrication \cite{Zighem_JAP2013}.
However, in first approximation, if this non linear and hysteretic
variation of $\delta H_{R}$ is neglected and adjusted by a linear
fit, an effective magnetoelectric coupling $\alpha_{ME}$ (in V.cm$^{-1}$.Oe$^{-1}$
being given the width of the piezoelectric actuator ($0.7$ cm)) can
be estimated. The linear adjustments to the experimental data are
presented in Figure \ref{Fig_Resonance_Field_Shift}. A strong magnetoelectric
coupling is found for the Ni/Kapton\textregistered{}/Piezoelectric
system with an effective magnetoelectric coefficient $\alpha_{ME}\sim0.2$
V.cm$^{-1}$.Oe$^{-1}$. Interestingly, in the case of the Ni/Si/Piezoelectric
system, it can be conclude that the magnetoelectric coupling is roughly
seven times less efficient because of its higher magnetoelectric coefficient
value ($\alpha_{ME}\sim1.35$ V.cm$^{-1}$.Oe$^{-1}$). The relatively
weak coupling found in the second system is due to a weaker transmission
of the in-plane elastic stresses. The strain loss of a factor of almost
7 cannot be explained only by the stiffness of Silicon but also by
the often reported imperfection of cementation of the epoxy glue\cite{Brandlmaier2008_PRB_Bis}.
Obviously these imperfections are insignificant when the substrate
is highly stretchable like polymers.

\begin{figure}
\includegraphics[bb=65bp 180bp 595bp 595bp,clip,width=8cm]{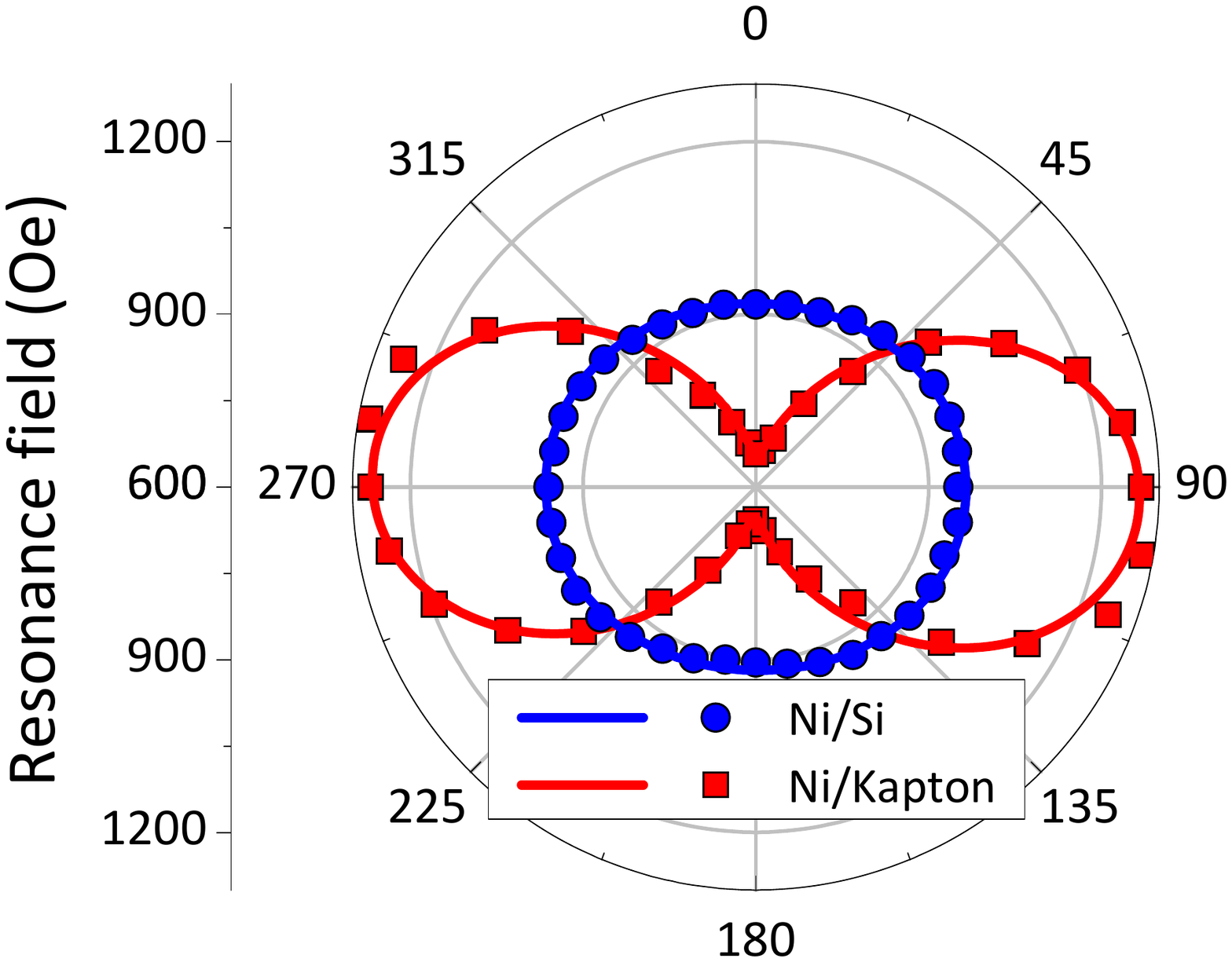}

\caption{In-plane angular dependence of the resonance field at zero applied
voltage for the two studied systems. Symbols refer to experimental
data while lines are best fits to the experimental data by using equations
\ref{eq:Hres-H1} and \ref{eq:Hres-H2}. These fits allow an evaluation
of the in-plane non-equibiaxial stress: $\left|\sigma_{xx}^{residual}-\sigma_{yy}^{residual}\right|=200$
MPa and $\left|\sigma_{xx}^{residual}-\sigma_{yy}^{residual}\right|=15$
MPa for the structures with a rigid and flexible substrate, respectively. }

\label{Fig_Residual_Stresses}
\end{figure}

However, depositing a magnetic thin film on a flexible substrate can
lead to an \textit{a priori} undesirable effects related to its possible
anisotropic non-flatness during elaboration process. Moreover, because
of the large contrast between the polymer and the Ni Young's modulus
($\sim$4 GPa and $\sim$200 GPa, respectively), a more or less pronounced
additional curvature due to growth stress develops during film deposition.
These phenomenona lead to a stress state in the magnetic film with
a possible in-plane non-equibiaxiallity\cite{Zhang2013}. This behavior
is illustrated in Figure \ref{Fig_Residual_Stresses} showing in-plane
angular dependencies (thereafter $\varphi_{H}$ is the angle between
the in-plane applied magnetic field and $x$-axis) of the resonance
field of the as-deposited structures. A huge uniaxial anisotropy (anisotropy
field close to 300 Oe) characterized by a horizontal peanut shape
of the resonance field variation is observed when the Ni film grown
on Kapton\textregistered{} whereas a very weak one (anisotropy field
around 30 Oe) is found when using the Si substrate. Since the present
Ni films are polycrystalline with no preferential crystallographic
orientations, no in-plane macroscopic magnetocrystalline effects are
expected in both cases. Thus this uniaxial anisotropy should have
a magnetoelastic origin. The influence of the residual stresses on
the magnetic properties of the as deposited films can be modeled using
an isotropic magnetoelastic anisotropy energy term $F_{ME}$:

\begin{equation}
F_{ME}=-\frac{3}{2}\lambda\left(\left(\gamma_{x}^{2}-\frac{1}{3}\right)\sigma_{xx}^{residual}+\left(\gamma_{y}^{2}-\frac{1}{3}\right)\sigma_{yy}^{residual}\right)
\end{equation}

$\sigma_{xx}^{residual}$ and $\sigma_{yy}^{residual}$ being the
in-plane principal residual stress tensor components while $\gamma_{x}$
and $\gamma_{y}$ correspond to the direction cosines of the in-plane
magnetization. In this condition the resonance field expression can
be written as $H_{R}=H_{1}+H_{2}$ with:

\begin{multline}
H_{1}=\Bigg[\Big(2\pi M_{s}+\frac{3\lambda}{2M_{S}}\big(\sigma_{xx}^{residual}\sin^{2}\varphi_{H}\\
+\sigma_{yy}^{residual}\cos^{2}\varphi_{H}\big)\Big)^{2}+\Big(\frac{2\pi f}{\gamma}\Big)^{2}\Bigg]^{0.5}-2\pi M_{S}\label{eq:Hres-H1}
\end{multline}

\begin{multline}
H_{2}=-\frac{3\lambda}{4M_{S}}\Bigg[\sigma_{xx}^{residual}(1+3\cos2\varphi_{H})\\
+\sigma_{yy}^{residual}(1-3\cos2\varphi_{H})\Bigg]\label{eq:Hres-H2}
\end{multline}

In the above expressions $H_{1}$ essentially represents a constant
shift in the resonance field baseline because $2\pi M_{S}$ ($M_{S}$
being the saturation magnetization) and $\frac{2\pi f}{\gamma}$ ($\gamma$
is the gyromagnetic factor) are found to be larger than the equivalent
magnetoelastic anisotropy field. The influence of in-plane residual
stresses on the magnetic properties is included in $H_{2}$ term.
It should be noted that this expression is obtained in the assumption
of a uniform magnetization (macrospin approximation) aligned along
the applied magnetic field. This assumption is well fulfilled at the
8 GHz driven frequency chosen for this study. It clearly appears that
an equibiaxial residual stress ($\sigma_{xx}^{residual}=\sigma_{yy}^{residual}$)
leads to an isotropic variation of the resonance field with a slight
increase or decrease of the mean value depending on the sign of the
residual stress. However, a non equibiaxial residual stress ($\sigma_{xx}^{residual}\neq\sigma_{yy}^{residual}$)
leads to an anisotropic angular variation of the resonance field.
Thanks to this model, the experimental angular variations of the resonance
field have been fitted (see continuous lines in Figure \ref{Fig_Residual_Stresses})
by using usual Ni material magnetic parameters: $M_{S}=480$ emu.cm$^{-3}$,
$\gamma=1.885\times10^{7}$ Hz.Oe$^{-1}$ and $\lambda=-26\times10^{-6}$\cite{Zighem_JAP2013}.
The deduced non-equibiaxiallity is $\left|\sigma_{xx}^{residual}-\sigma_{yy}^{residual}\right|=200$
MPa and $\left|\sigma_{xx}^{residual}-\sigma_{yy}^{residual}\right|=15$
MPa for the Ni film deposited onto flexible and rigid substrate, respectively.
These results show that MS-FMR experiments can also be used for the
determination of the non-equibiaxial residual stress in magnetostrictive
films. It should be noted here that this kind of knowledge is not
straightforward to get with standard techniques (sample deflection,
X-ray diffraction), for which an equibiaxial stress state is generally
assumed. Indeed, this parameter cannot be neglected since it determines
the initial magnetization direction in the magnetic thin film being
considered.

In conclusion, an optimization of the indirect magnetoelectric coupling
in thin-film/substrate/piezoelectric-actuator heterostructure has
been performed by employing a polymer as a substrate. This optimization
is due to a better strain transmission between the piezoelectric actuator
and the Ni film which is due to the weak Young's modulus of the Kapton\textregistered{}.
Furthermore, at zero applied voltage, a huge magnetoelastic anisotropy
is evidenced in the Ni/Kapton\textregistered{}/Piezoelectric system.
It is attributed to a {}``residual'' non-equibiaxial stress state
due to a slight curvature along a given direction which generally
appears when depositing a metallic film onto a flexible substrate.
However, this effect could be a limitation for some applications where
weak in-plane magnetic anisotropies are required.
\begin{acknowledgments}
The authors gratefully acknowledge the CNRS for his financial support
through the {}``PEPS INSIS'' program (FERROFLEX project). This work
has been also partially supported by the Université Paris XIII through
a {}``Bonus Qualité Recherche'' project. \end{acknowledgments}

\end{document}